\documentclass[twocolumn]{article}
\usepackage{times}
\include{psfig}
\usepackage{epsfig}
\usepackage{amssymb}
\begin{document}
\title{Strong correlations between text quality and complex networks features}

\author{
L. Antiqueira$^1$, M.G.V. Nunes$^1$, O.N. Oliveira Jr.$^2$ and 
L. da F. Costa$^2$ \\
1 -- Instituto de Ci\^encias Matem\'aticas e de Computa\c{c}\~ao, \\
Universidade de S\~ao Paulo, CP~668, \\
13560-970, S\~ao Carlos, SP, Brazil \\
2 -- Instituto de F\'{\i}sica de S\~ao Carlos, \\
Universidade de S\~ao Paulo, \\
CP~369, 13560-970, S\~ao Carlos, SP, Brazil
}

%
%
%

\maketitle

\emph{Abstract:}
Concepts of complex networks have been used to obtain metrics that
were correlated to text quality established by scores assigned by
human judges. Texts produced by high-school students in Portuguese
were represented as scale-free networks (word adjacency model), from
which typical network features such as the in/outdegree, clustering
coefficient and shortest path were obtained. Another metric was
derived from the dynamics of the network growth, based on the
variation of the number of connected components. The scores assigned
by the human judges according to three text quality criteria
(coherence and cohesion, adherence to standard writing conventions and
theme adequacy/development) were correlated with the network
measurements. Text quality for all three criteria was found to decrease
with increasing average values of outdegrees, clustering coefficient
and deviation from the dynamics of network growth. Among the criteria
employed, cohesion and coherence showed the strongest correlation,
which probably indicates that the network measurements are able to
capture how the text is developed in terms of the concepts represented
by the nodes in the networks. Though based on a particular set of
texts and specific language, the results presented here point to
potential applications in other instances of text analysis.

\section{Introduction}
\label{sec:intro}

The area of complex networks
\cite{Albert2002,Dorogovtsev2002,Newman2003}, which can be viewed as
an intersection between graph theory and statistical mechanics, has
been marked by many theoretical advances and relevant applications
over the last few years. New concepts such as the hubs, i.e. nodes
with particularly high degree, had major impact for understanding and
re-interpreting problems such as essentiality \cite{Jeong2001} and
resilience to attacks \cite{Schwartz2002}. Applications of complex
networks have appeared in widely diverse areas, ranging from the
Internet \cite{Albert1999} to networks of Jazz artists
\cite{Gleiser2003}. Because of its special importance to human
communication, culture, and even intelligence, the representation and
analysis of written texts in terms of graphs and complex networks
offers a promising (and challenging) research opportunity for the
forthcoming years. The application of concepts and tools from
mathematics, physics and computer science to the analysis of texts is
not new and includes approaches generally associated with first-order
statistics of words and other elements obtained from texts. With the
availability of databases accessible through the Internet,
unprecedented possibilities for such investigations are now open. For
instance, considering first-order statistics, Gon\c{c}alves \& Gon\c{c}alves
have identified the works of renowned English writers
\cite{Goncalves2005}, Montemurro \& Zanette have grouped words based
on their linguistic role in a corpus \cite{Montemurro2002}, and Zhou
\& Slater have proposed a method to measure the relevance of words in
a text \cite{Zhou2003}. Indeed, first-order analysis does provide
valuable information about the global and specific features of most
texts.

We believe that further insights can be obtained by using higher-order
statistics in order to enhance the context representation, to which
the concept of complex networks is closely related. More specifically,
each word in a text can be represented as a node, while subsequent
words define associations, or edges, between such nodes (this model is
known as word adjacency/co-ocurrence network). Typically, pairs of
subsequent words, excluding articles and other connecting words, are
considered, implying a Markov model with unity length memory. Larger
Markov memory lengths can be obtained by considering tuples of
subsequent words. Because the networks incorporate the most immediate
associations between words and concepts, their topology - quantified
by several measurements~\cite{Costa2005a} such as node degree,
clustering coefficient and shortest path - can provide information on
some properties of the text, such as style and authorship. A series of
studies indicate that word adjacency networks
\cite{Cancho2001,Dorogovtsev2001,Allegrini2003,Milo2004}, semantic
networks
\cite{Albert2002,Steyvers2001,Kinouchi2001,Sigman2002,Motter2002,Holanda2004,Dorow2005},
word association networks \cite{Steyvers2001,Costa2004,Capocci2004}
and syntactic networks \cite{Cancho2004,Cancho2005} are graphs that
show features present in classical examples of complex networks, such
as the World Wide Web and social networks. One of the important
consequences of such studies is the presence of hubs in linguistic
networks.

In this study we investigate the possibility of automated evaluation
of text quality using topological measurements extracted from the
corresponding complex networks. We consider three criteria for scoring
texts which are related to text quality, namely i) coherence and
cohesion, (ii) adherence to standard writing conventions and (iii)
theme adequacy/development. These are the criteria generally employed
to mark essays for high-school students applying to enter the
university in Brazil. Complex networks are obtained from such texts by
considering the proximity between words, and the indegree and
outdegree, the clustering coefficient and shortest path distributions
are estimated for each text. Such measurements are estimated after the
full construction of the networks, while the number of connected
components is monitored during their growth, yielding a topological
feature which is a function of the number of added word
associations. All the measurements are correlated with grades assigned
by human experts. The results indicate that, despite the many
parameters and unavoidable subjectivity of human language, such an
approach presents potential to be used as a subsidy for a more
objective and reproducible means to evaluate text quality.


\section{Text assessment by human subjects}
\label{sec:assess}

\begin{figure}
	\centering
	\resizebox{0.9\columnwidth}{!}{
		\includegraphics{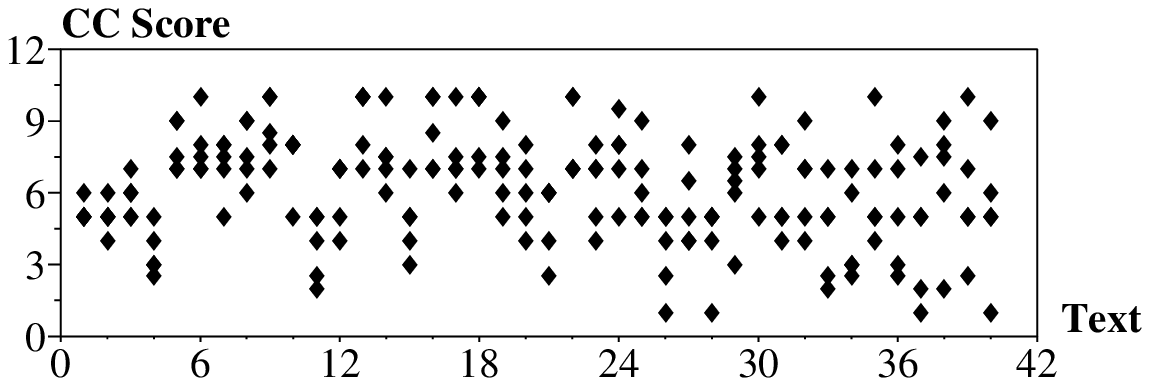}
	}
	\resizebox{0.9\columnwidth}{!}{
		\includegraphics{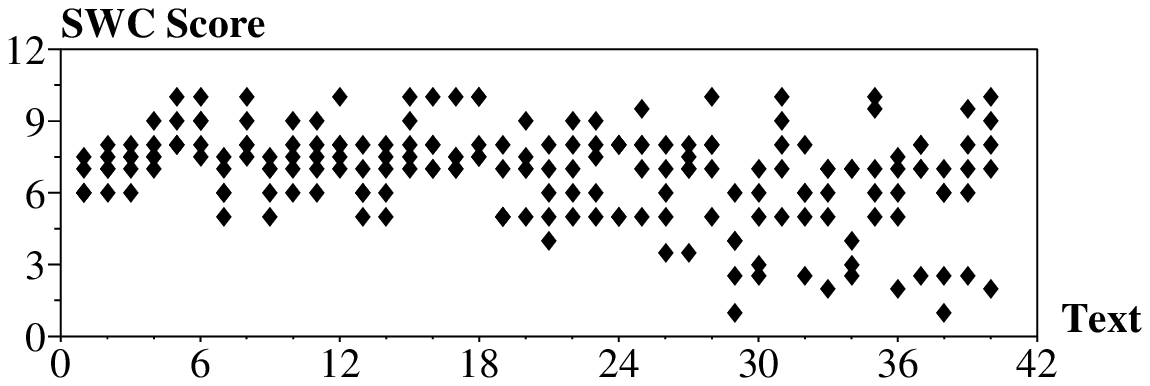}
	}
	\resizebox{0.9\columnwidth}{!}{
		\includegraphics{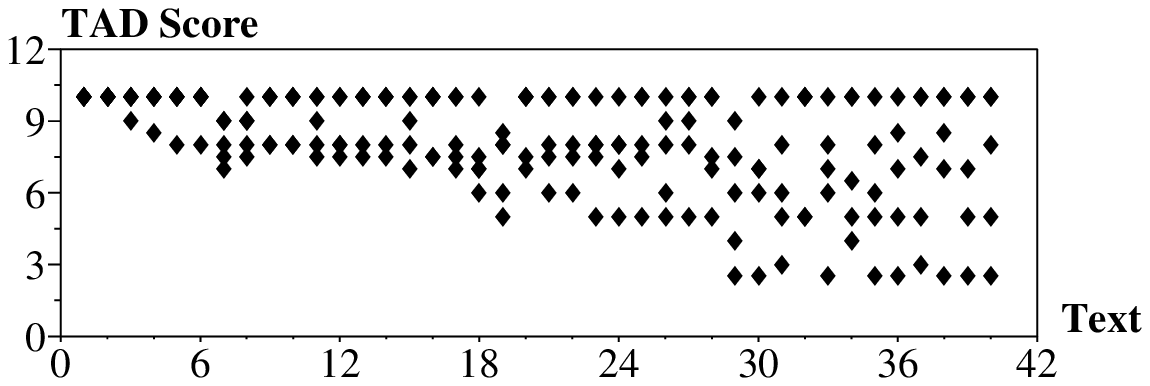}
	}
	\caption{40 texts vs. corresponding scores (from 0 to 10)
	according to the three quality criteria, identified as $CC$
	for coherence and cohesion, $SWC$ for adherence to standard
	writing conventions and $TAD$ for theme
	adequacy/development. The horizontal axes are ordered from the
	text with the lowest score dispersion between the five judges
	to the text with the highest dispersion. The sequences of
	texts in the horizontal axes are not necessarily equal for the
	different criteria.}
	\label{fig:scores}
\end{figure}

One set of 40 pieces of text has been selected, which comprise essays
on the same subject - influence from TV on our society - written in
Brazilian Portuguese by high school students. All pieces of text have
approximately the same size, with an average of 228 words. A panel of
five human judges, all of which are computational linguists, analyzed
the texts using three criteria to mark them, namely (i) coherence and
cohesion, (ii) adherence to standard writing conventions and (iii)
theme adequacy/development, henceforth referred to as $CC$, $SWC$ and
$TAD$, respectively. The judges assigned marks from 0 to 10 to each
text for the three criteria, and did not receive any instruction as to
reference values or how these criteria should be rated. Not
surprisingly, there is large dispersion among the marks given, as
illustrated in Fig.~\ref{fig:scores}, where the five marks are shown
in the vertical axes for each of the 40 texts (horizontal axes). The
texts were sorted from left to right according to an increasing
dispersion in the scores assigned by the judges. The numbering of the
text may therefore vary from one figure to the other, as the different
criteria were analyzed. Note also that for some texts less than five
points may appear in the picture because equal scores could have been
given. Because of the large dispersion of the marks, in this paper we
shall concentrate on data obtained with the 20 texts with lowest score
dispersion. The results obtained with the full set of 40 texts will be
briefly commented upon in the Concluding section.

\section{Measurements of text features using complex networks}
\label{sec:measur}

Two word adjacency networks were obtained from a given text. In the
first one, called \mbox{NET-A}, each different pair $(w_1,w_2)$ of subsequent
words (at distance one from each other) defines a directed weighted
edge in the network, whose weight represents the frequency of the
association from word $w_1$ to word $w_2$. The association network was
obtained similarly to that described in \cite{Costa2004}, i.e. each of
the $N$ different words was represented as a node and each connection
between words as a weighted edge between the corresponding nodes
representing those words. The stopwords have been removed and the
remaining words have been lemmatized. Removing stopwords eliminates
very common words, such as verb to be and some adverbs, and words from
closed classes (articles, pronouns, prepositions and
conjunctions). Lemmatization is the process of reducing a word into
its base form, such as the verb ``passed'' to the infinitive
``pass''. Therefore, different occurrences of meaning-related words are
represented by the same node in the network. The second word adjacency
network, called \mbox{NET-B}, is almost the same as \mbox{NET-A}, but also connects
subsequent words at distance two, i.e. $w_1$ is also connected to
$w_3$, although there is a word $w_2$ between them.  In other words,
each sequence of three words $(\ldots, w_1, w_2, w_3, \ldots)$ implies
a directed edge from $w_1$ to $w_3$ and another directed edge from
$w_2$ to $w_3$.  Note that the two adopted types of networks, namely
\mbox{NET-A} and \mbox{NET-B}, represent Markov models of memory one and two,
respectively.  The choice of these two models has been aimed at
providing subsidies for investigating the effect of the extent of the
considered context into the measurements and results.

All network measurements adopted were extracted from the weight matrix
$W$ representing the network. This $N \times N$ matrix was obtained by
starting with all elements as zero and making $W(j,i)=W(j,i)+1$
whenever there was the association $i \rightarrow j$. Because of the
directed edges, the matrix $W$ is not symmetric. It is also possible
to obtain an adjacency matrix $K$ from $W$ by making $K(j,i)=1$
whenever $W(j,i) > 0$. The measurements obtained from such networks
are described in the remaining of this section.

\subsection{Indegree and outdegree}

The indegree and outdegree of node $i$ are defined, respectively, as

\begin{equation}
ID(i)=\sum^{N}_{j=1}{W(i,j)} \label{eq:indegree}
\end{equation}

and

\begin{equation} OD(i)=\sum^{N}_{j=1}{W(j,i)} \label{eq:outdegree} .
\end{equation}
We adopt the network outdegree $OD$ as the arithmetic mean of every
$OD(i)$ (the network indegree $ID$ is obtained similarly). Because the
average value of the indegrees coincides with that obtained for the
outdegrees, only the latter will be considered henceforth.

\subsection{Clustering coefficient}

\begin{figure}
	\centering
	\resizebox{0.8\columnwidth}{!}{
    	\includegraphics{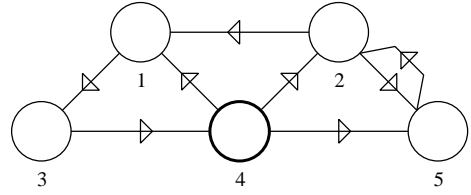}
    }
	\caption{Computation of the clustering coefficient of node~4
	($CLC(4)$). In this particular case, $N_c = 3$, since node~4
	is connected to the nodes belonging to the set $R = \{1,2,5\}$
	(node~3 has an edge shared with node~4, but this edge does not
	come from node~4). If the nodes 1, 2 and 5 formed a fully
	connected subnetwork, there would be $N_c(N_c-1) = 3(3-1) = 6$
	edges between them, but in fact there is only $B =
	3$. Finally, the clustering coefficient of node~4 is $CLC(4) =
	B / (N_c(N_c-1)) = 3/6 = 0.5$. This definition of clustering
	coefficient does not take into account the edge weights.}
	\label{fig:clc}
\end{figure}

The clustering coefficient of node $i$ is calculated as
follows. First, all nodes receiving an edge from node $i$ are
identified and included into the set $R$, with $N_c = |R|$. If $B$ is
the total number of edges between all the nodes in $R$ (taking into
account the edges directions, i.e. edge $i \rightarrow j$ is different
from edge $j \rightarrow i$), the clustering coefficient of node $i$
is obtained as (for an example, see Fig.~\ref{fig:clc})
\begin{equation} CLC(i) = \frac{B}{N_c(N_c-1)} . \label{eq:clustcoeff}
\end{equation} In case $N_c$ is smaller or equal to 1, then $CLC(i) =
0$. The network clustering coefficient $CLC$ is the arithmetic mean of
all individual clustering coefficients $CLC(i)$.

\subsection{Network dynamics}

\begin{figure*}
	\centering
	\resizebox{1.8\columnwidth}{!}{
		\includegraphics{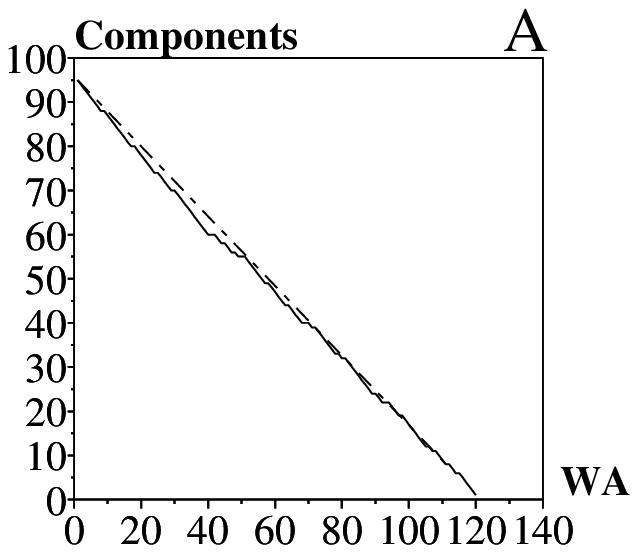}
		\includegraphics{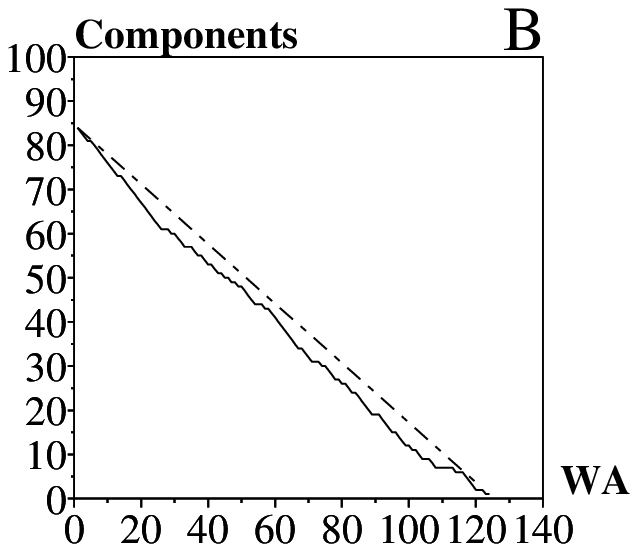}
		\includegraphics{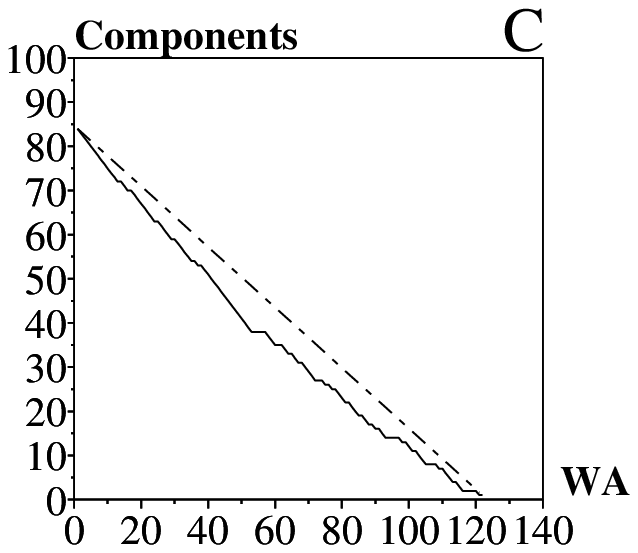}
	}
	\caption{Dynamics of the network for three texts extracted
	from the selected set of 40 pieces of text. In the horizontal
	axes, $WA$ stands for the number of word associations already
	inserted into the network. The straight dotted line is a
	reference that assumes uniform variation of the number of
	components as the edges are inserted or as their corresponding
	weights are modified in the network. The other curve is the
	real one, which reflects the actual variation of the number of
	components. The deviation in the network dynamics, according
	to Equation~\ref{eq:compdyn}, for the three texts above are
	0.014 (A), 0.045 (B) and 0.064 (C).}  \label{fig:components}
\end{figure*}

We have taken measurements considering the dynamics of growth for the
complex network as a given text was analyzed. The number of connected
components (or clusters) was calculated after adding each word
association to the network, yielding a topological feature which is a
function of the number of associations and, consequently, of the
evolution of the text construction. For each text, the network was
initiated with all $N$ words, each one representing a single
component, and the connections were established by each word
association that occurred along the text. When a word association was
read, a new edge was created in the network or the weight of an
already existing edge was increased. As a consequence of the word
adjacency model, the number of connected components always converged
to one after all words had been introduced. Fig.~\ref{fig:components}
shows how the number of components evolves with the number of edges
for three texts extracted from the selected corpus, being therefore
representative of the evolution of connectivity in a given text. In
each graph, the straight line was included to guide the eye, and
represents the special case of uniform variation of the number of
components, while the other curve indicates the real variation as the
word associations were read. A quantitative treatment of the data in
Fig.~\ref{fig:components} was carried out by calculating the extent to
which the real plot departed from the straight line. For short, this
measurement will be referred to in the remainder of this article as
``components dynamics deviation'' ($CDD$).  Let $f_a(x)$ be the actual
function that associates the number of components with the number $x$
of word associations already inserted into the network, $f_s(x)$ be
the reference straight line, $L$ be the total number of word
associations in the text and $N$ be the total number of vertices in
the network. The deviation in the network dynamics is calculated as
\begin{equation} CDD = \frac{\sum_{x=1}^{L}{|f_a(x)-f_s(x)|/N}}{L}
\label{eq:compdyn} . \end{equation} Texts A, B and C, whose dynamics
are represented in Fig.~\ref{fig:components}, have $CDD$ values of
0.014, 0.045 and 0.064, respectively. A visual inspection of these
three texts in Fig.~\ref{fig:components} corroborates these increasing
values obtained for texts from A to C.

\subsection{Shortest path}

Distances between pairs of nodes, which also consider the edges
directions, were calculated with the Floyd-Warshall algorithm
\cite{Cormen2001}. We consider the complement of each weight,
$W_{max}-W(j,i)+1$, where $W_{max}$ is the maximum edge weight present
in the network, to compute the shortest paths $SP(i,j)$ between any
two nodes $i$ and $j$. $SP$ is defined in this way because its is not
desirable that the shortest paths algorithm gives low priority to the
strongest edges, which are those that represent more frequent and
possibly more important associations between words. Whenever there is
no path between two nodes $i$ and $j$, we take $SP(i,j) = N W_{mean}$,
where $N$ is the number of vertices and $W_{mean}$ is the arithmetic
mean of all edge weights. The $SP$ measurement for a whole network is
the arithmetic mean of every $SP(i,j)$, provided that $i \neq j$.


\section{Results and discussion}
\label{sec:results}

\begin{figure*}
	\centering
	\resizebox{1.8\columnwidth}{!}{
		\includegraphics{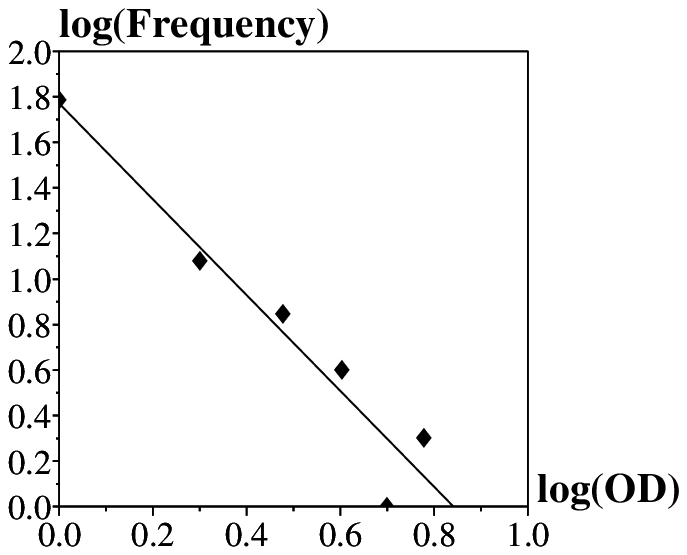}
		\includegraphics{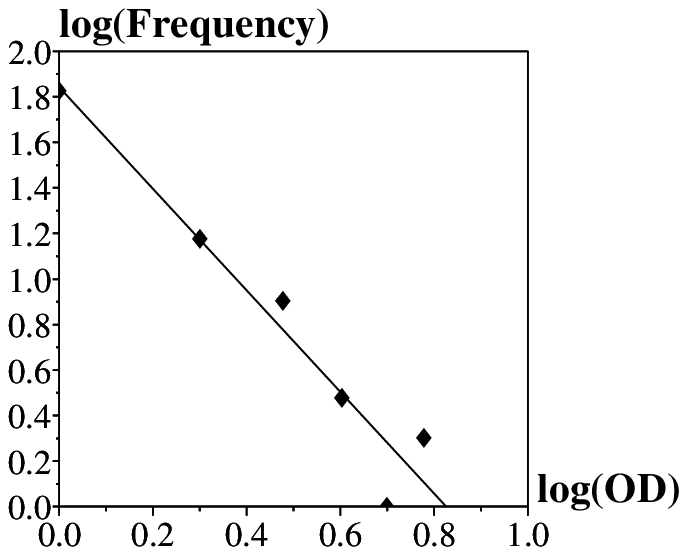}
		\includegraphics{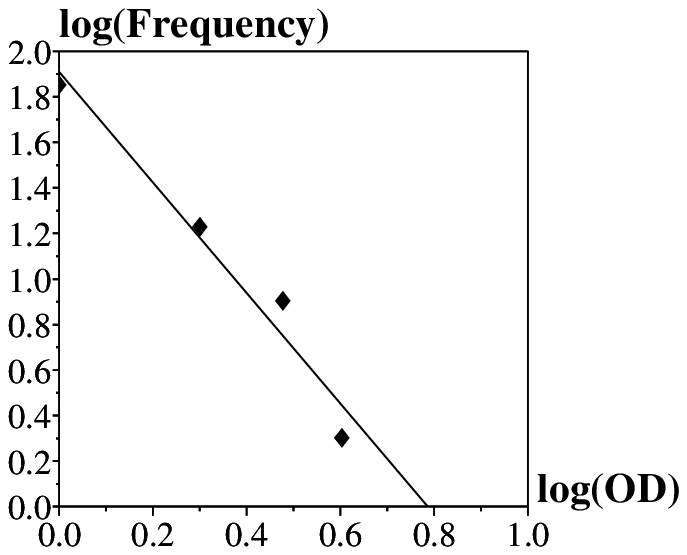}
	}
	\caption{Log$\times$log outdegree ($OD$) distributions for
	three texts extracted from the corpus of 40 texts. A
	scale-free behavior is suggested by these examples.}
	\label{fig:scale-free}
\end{figure*}

In a previous report \cite{Antiqueira2005a}, we have shown that the
measurements associated with complex networks could be used to
distinguish between low-quality and high-quality texts, selected from
two different sources. However, a limitation to that study was that
the differences emerging from the analysis could arise from the source
of the text, age and background of the writers and even subject of the
essays. In order to avoid such possible interferences, in the present
study we took texts from only one source, namely essays written by
high-school students, with approximately the same age and academic
background, on a single topic - influence from TV on our
society. Firstly, we illustrate in Fig.~\ref{fig:scale-free} for three
texts from the set that the distribution of outdegrees of the
investigated data suggest the scale-free property, indicated by the
linear log$\times$log plot for the outdegree, which is consistent with
previous reports in the literature
\cite{Cancho2001,Dorogovtsev2001}. Similar results were obtained for
the indegree and for the other texts (not shown here).

\begin{figure*}
	\centering
	\resizebox{1.7\columnwidth}{!}{
		\includegraphics{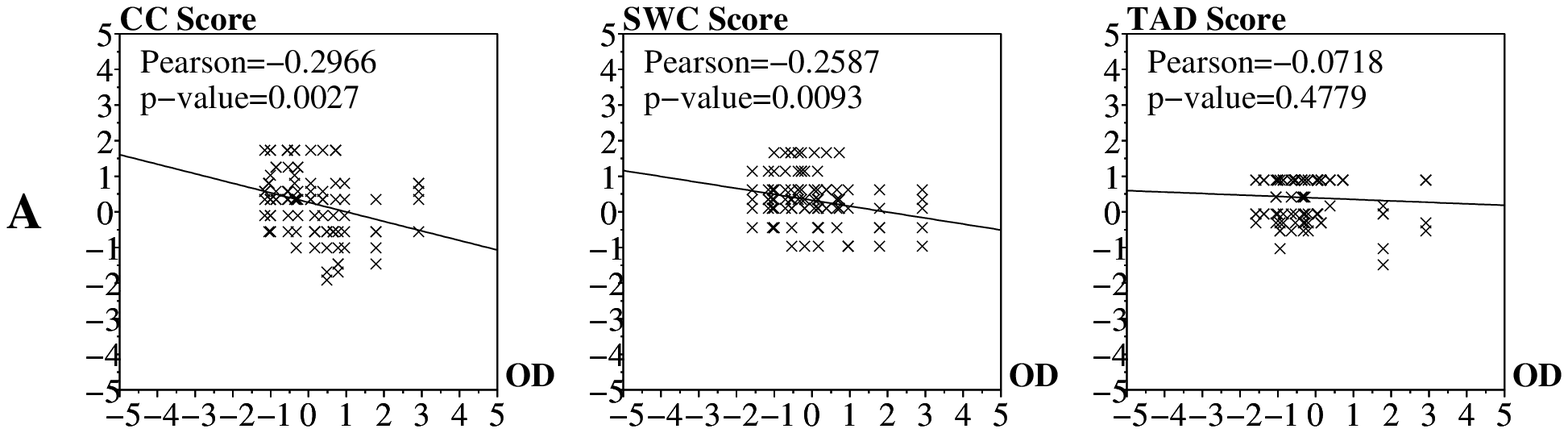}
	}
	\resizebox{1.7\columnwidth}{!}{
		\includegraphics{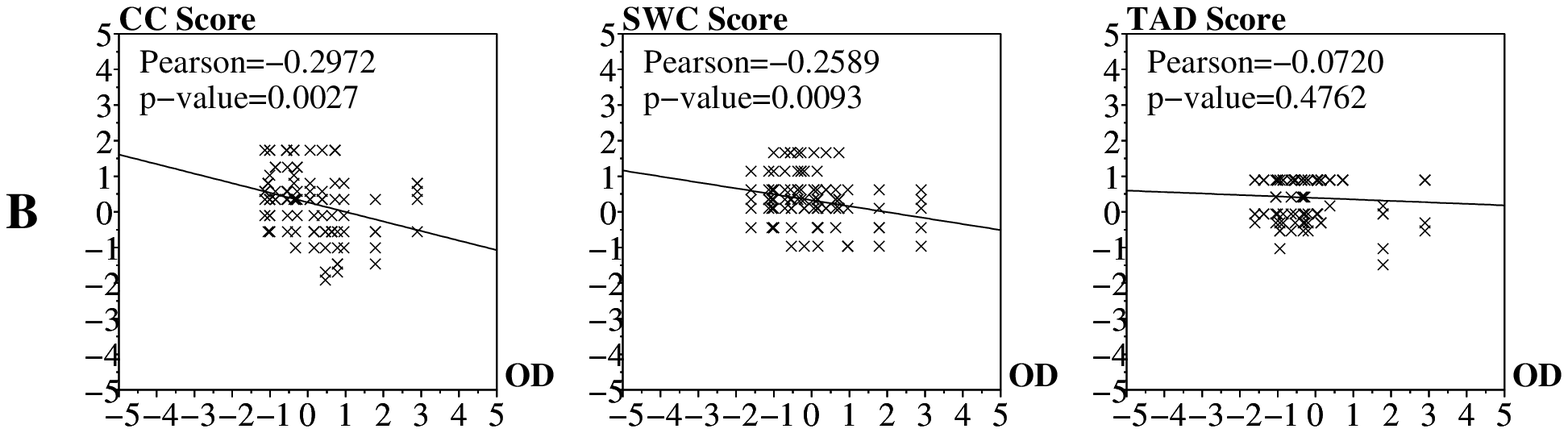}
	}
	\caption{Correlations between the outdegrees ($OD$, horizontal
	axes) and the scores (vertical axes) for the 20 texts with the
	lowest score dispersion. In the vertical axes, $SWC$ stands
	for standard writing conventions, $TAD$ for theme
	adequacy/development and $CC$ for coherence and cohesion. Both
	axes are standardized into a standard normal distribution
	$N(0,1)$. Measurements obtained from the two types of
	networks, \mbox{NET-A} and \mbox{NET-B}, are discriminated by the labels A
	and B, respectively.}
	\label{fig:corr-od}
\end{figure*}

\begin{figure*}
	\centering
	\resizebox{1.7\columnwidth}{!}{
		\includegraphics{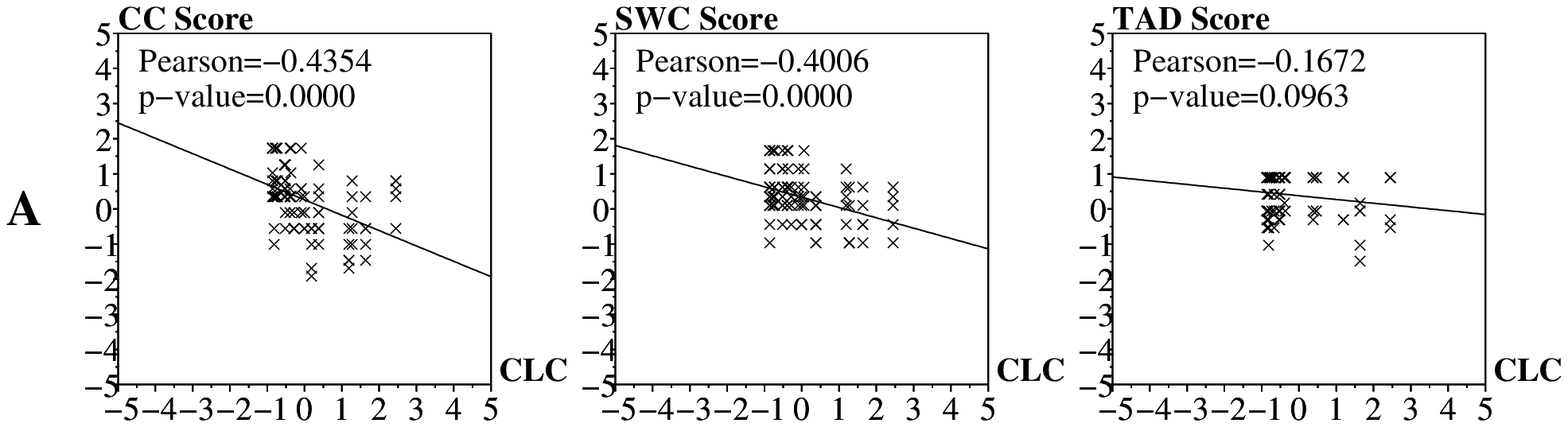}
	}
	\resizebox{1.7\columnwidth}{!}{
		\includegraphics{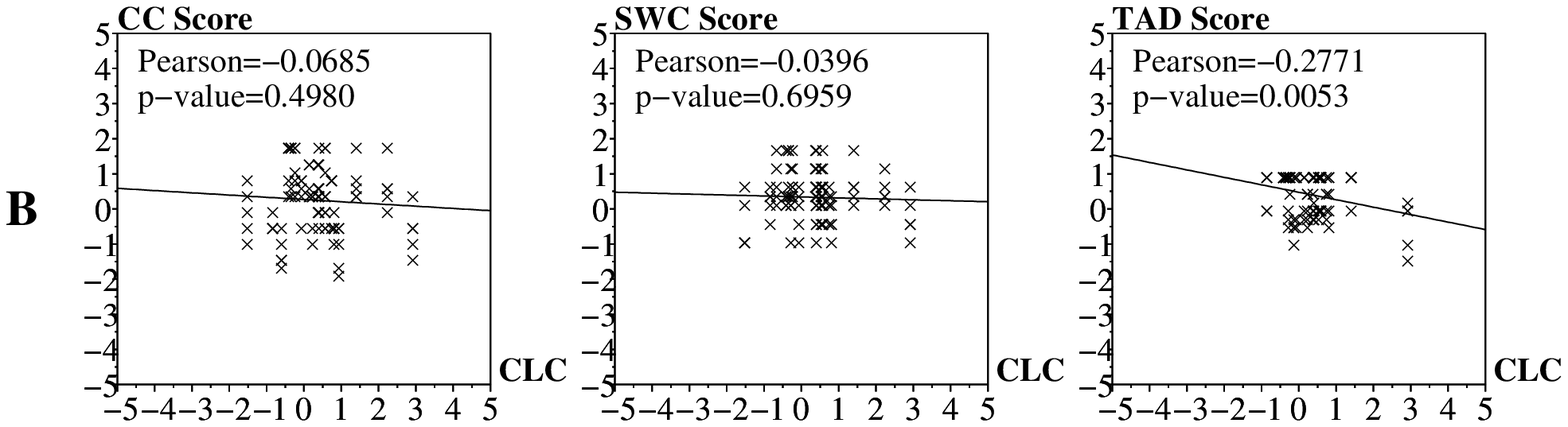}
	}
	\caption{Correlations between the clustering coefficients
	($CLC$, horizontal axes) and the scores (vertical axes) for the
	20 texts with the lowest score dispersion. In the vertical
	axes, $SWC$ stands for standard writing conventions, $TAD$ for
	theme adequacy/development and $CC$ for coherence and
	cohesion. Both axes are standardized into a standard normal
	distribution $N(0,1)$. Measurements obtained from the two
	types of networks, \mbox{NET-A} and \mbox{NET-B}, are discriminated by the
	labels A and B, respectively.}
	\label{fig:corr-clc}
\end{figure*}

\begin{figure*}
	\centering
	\resizebox{1.7\columnwidth}{!}{
		\includegraphics{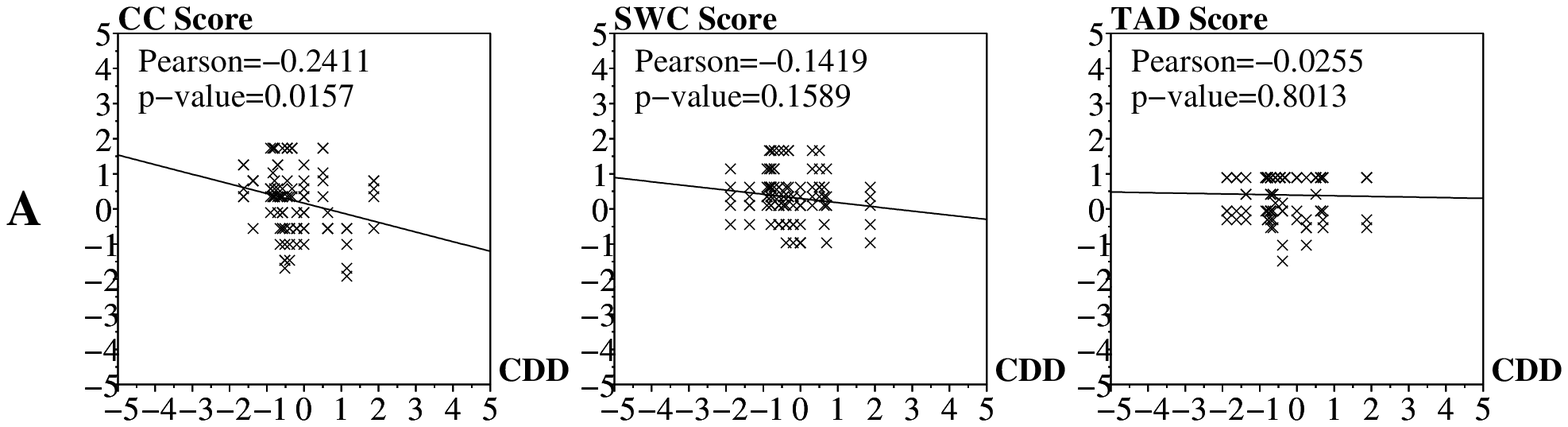}
	}
	\resizebox{1.7\columnwidth}{!}{
		\includegraphics{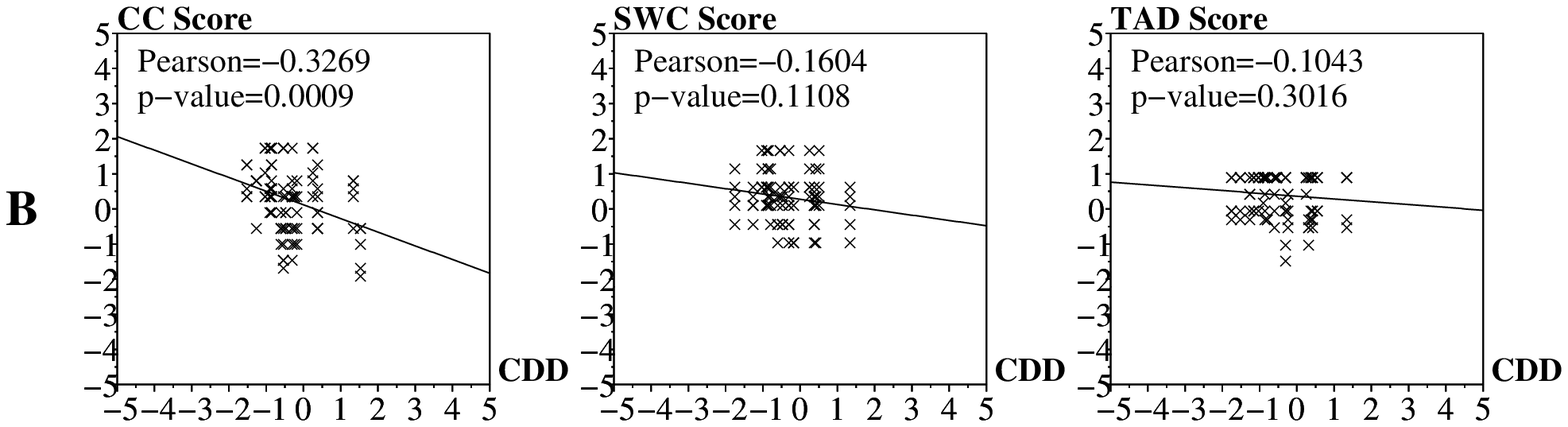}
	}
	\caption{Correlations between the components dynamics
	deviations ($CDD$, horizontal axes) and the scores (vertical
	axes) for the 20 texts with the lowest score dispersion. In
	the vertical axes, $SWC$ stands for standard writing
	conventions, $TAD$ for theme adequacy/development and $CC$ for
	coherence and cohesion. Both axes are standardized into a
	standard normal distribution $N(0,1)$. Measurements obtained
	from the two types of networks, \mbox{NET-A} and \mbox{NET-B}, are
	discriminated by the labels A and B, respectively.}
	\label{fig:corr-cdd}
\end{figure*}

\begin{figure*}
	\centering
	\resizebox{1.7\columnwidth}{!}{
		\includegraphics{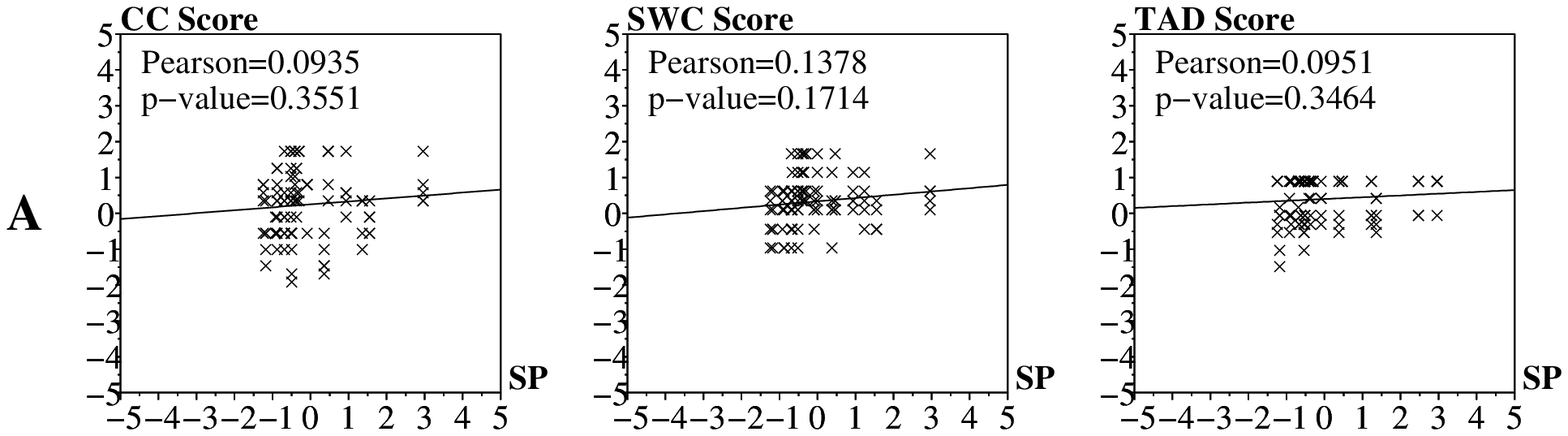}
	}
	\centering
	\resizebox{1.7\columnwidth}{!}{
		\includegraphics{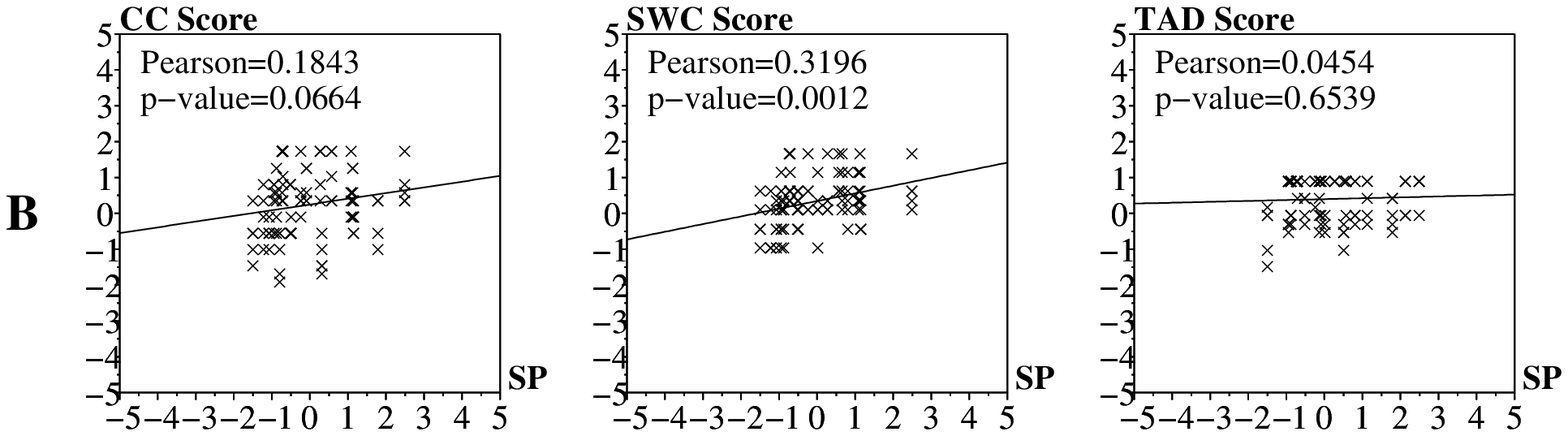}
	}
	\caption{Correlations between the shortest paths ($SP$,
	horizontal axes) and the scores (vertical axes) for the 20
	texts with the lowest score dispersion. In the vertical axes,
	$SWC$ stands for standard writing conventions, $TAD$ for theme
	adequacy/development and $CC$ for coherence and cohesion. Both
	axes are standardized into a standard normal distribution
	$N(0,1)$. Measurements obtained from the two types of
	networks, \mbox{NET-A} and \mbox{NET-B}, are discriminated by the labels A
	and B, respectively.}
	\label{fig:corr-sp}
\end{figure*}

We now attempt to correlate the measurements using complex networks
concepts with the scores assigned by the human judges. Because of the
large score dispersion for some texts, we perform the analysis taking
only the 20 texts with the lowest dispersion for each criterion. This
analysis results in a set of 24 plots
(Figs.~\ref{fig:corr-od}--\ref{fig:corr-sp}) which correlate the four
network measurements with the three types of scores for each of the
two types of networks. Figs.~\ref{fig:corr-od}--\ref{fig:corr-sp} are
organized with the measurements distributed along the horizontal axes,
while the scores assigned by the human judges are positioned in the
vertical axes. The labels A and B refer to the measurements taken from
the networks constructed following the models \mbox{NET-A} and \mbox{NET-B},
respectively. The values from both the measurements and scores were
standardized into a standard normal distribution $N(0,1)$ and a linear
regression was performed for each correlation plot. The corresponding
straight line, the Pearson correlation coefficient and the p-value are
also given in the figure as a guide for the strength of the linear
correlations \cite{Neter1996}.

Figs.~\ref{fig:corr-od}A and \ref{fig:corr-od}B indicate that the
scores assigned by the human judges - for the three criteria -
decrease with increasing number of outdegrees. Most significant are
the results for the cohesion and coherence ($CC$) and adherence to
standard writing conventions ($SWC$). Large Pearson coefficients were
obtained with very low p-values, which indicates that the linear
correlations were not obtained by chance.  The scores corresponding to
the theme adequacy/development ($TAD$) are less sensitive to the number
of outdegrees, though they also tend to decrease. It appears then that
an analysis of the number of outdegrees allows one to capture the
quality of the text, with a large number of outdegrees causing the
text to lose quality. There is practically no difference in behavior
in the results using \mbox{NET-A} and \mbox{NET-B}, i.e. considering a larger
context in \mbox{NET-B} did not affect the results significantly. It should
be mentioned that we have also calculated the indegrees for all of the
texts separately. Because averages were taken, the results were
identical to those of the outdegrees and were therefore omitted.

Similar conclusions can be drawn from Figs.~\ref{fig:corr-clc}A and
\ref{fig:corr-clc}B, which show that text quality decreases with an
increasing clustering coefficient ($CLC$). Now correlations appeared
stronger for the data with \mbox{NET-A} than for \mbox{NET-B}, particularly for the
$CC$ and $SWC$ scores. In fact, from all measurements those of $CLC$ gave
the highest correlations (cf. Pearson coefficient) with text
quality. From a linguistic point of view, one may infer that texts
lose quality if the concepts are highly interconnected, probably
excessively interconnected.

As for the deviation from a linear dynamics for the network growth
($CDD$), an inspection of Figs.~\ref{fig:corr-cdd}A and
\ref{fig:corr-cdd}B points to the text quality decreasing with
increasing deviations, with little difference between data for \mbox{NET-A}
and \mbox{NET-B}. This corroborates our earlier finding with texts from two
different sources (see first version of this paper
\cite{Antiqueira2005a}). In the latter study, a threshold in the $CDD$
value was used to distinguish between low- and high-quality texts. A
large deviation indicates that the concepts were first introduced at
an early stage of the text construction, thus causing the total number
of components to decrease fast. As a result, the writer probably kept
repeating the arguments in the remainder of the writing process,
leading to a low quality text. As an example of this correlation,
consider the texts whose dynamics are illustrated in
Fig.~\ref{fig:components}. These texts received average scores of 7.9
(A), 5.2 (B) and 3.7 (C), according to the coherence and cohesion
criterion, while the $CDD$ values were 0.014 (A), 0.045 (B) and 0.064
(C), respectively. From a linguistic point of view, $CDD$ appears to
capture whether the flow of the prose is adequate, which is reflected
especially in the cohesion and coherence.

The correlation between the scores used to assess quality and the
measurements of shortest paths is weaker than for the other
measurements obtained with \mbox{NET-A} and \mbox{NET-B}, as shown in
Figs.~\ref{fig:corr-sp}A and \ref{fig:corr-sp}B. There is a slight
increase in the quality scores with increasing shortest paths,
especially with the $SWC$. The reason for a weaker correlation may be
found in the results from our previous work with texts of different
sources~\cite{Antiqueira2005a}. There, we found that text quality
appeared to increase slightly with the shortest path when all texts
were considered. However, when analyzing only the low-quality texts,
we observed text quality to decrease with increasing shortest
paths. We interpreted the latter result as being due to the
difficulties faced by poor writers in establishing long sequences of
connections among different concepts. This discrepancy between low and
high-quality texts calls for further, more detailed research into the
possible correlation between shortest paths and quality.

\section{Conclusions and perspectives}
\label{sec:concl}

We have applied the concepts of complex networks to one set of texts
which comprises essays of variable quality (as confirmed by human
judges) written by high-school students on the same topic. A
correlation could be established between the measurements,
i.e. outdegrees, clustering coefficient and deviation from a linear
dynamics in the network growth, and the scores assigned by the human
judges. The influence of shortest paths on text quality could not be
established unequivocally, probably because the effects may differ for
low and high-quality texts. Among the criteria employed, cohesion and
coherence was the one showing strongest correlation between the scores
and the network measurements. One may argue that this correlation
indicates that the measurements are able to capture how the text is
developed in terms of the concepts represented by the nodes in the
networks. We should not expect these measurements to capture the text quality in terms of the
adherence to standard writing conventions ($SWC$), as there is no deep
analysis of the texts. Essays performing well in this criterion were
those with small or negligible number of spelling and grammatical
mistakes. However, writers that produce good-quality texts in terms of
cohesion and coherence normally write grammatically correct texts. We
believe this to be the reason for the good correlation between the
scores of $SWC$ and the network measurements. The third criterion, theme
adequacy/development ($TAD$) is a more subjective because human judges
assess whether the writer addressed the expected issues for the given
topic. It is not uncommon that the score assigned be related to
whether the examiner agrees with the ideas put forward in the
essay. Not surprisingly then, the correlation between the network
measurements and the scores was weak.

The conclusions above hold for the 2 types of analysis performed, both
with \mbox{NET-A} and \mbox{NET-B}. Therefore, the context captured with only
adjacent words appears to be sufficient to correlate with text
quality. In addition, in subsidiary experiments we observed that
essentially the same conclusions and trends apply for the full set of
40 texts, which also included those with large dispersions in the
scores assigned by the human judges (results not shown here). The
trend toward decreasing scores with the number of outdegrees and
clustering coefficient suggests that text lose quality if the concepts
are highly interconnected. With the analysis of the network dynamics,
one infers that the faster and closer a writer introduces new concepts
not seen so far in a text, the worse the text is.

Though based on a particular set of texts and specific language, the
results presented here point to potential applications in other
instances of text analysis.  Indeed, the relatively high correlations
obtained between human assessment and network measurements are to some
extent surprising because of the potential complexity and subjectivity
underlying text quality and human language. One can now envisage, for
instance, an expert system that automatically marks essays, based on
machine learning methods. This will require golden standards, with a
panel of human judges agreeing on scores for a given set of texts (say
100 texts), with very little dispersion. If the network measurements
are taken for these manually-marked essays and associated with the
corresponding scores, machine learning algorithms may be used to
classify the remaining texts. This is an interesting scenario for
exams involving thousands of essays. Moreover, for essays with a
pre-defined specific topic the expert system could be further
sophisticated to consider the use of expected concepts and
associations among these concepts. Finally, the approach presented
here paves the way for the concepts of complex networks to be applied
to other types of text, as in the identification of text genres and
authorships, in addition to systems of information retrieval and
automatic summarization. This may have a large impact in areas such as
natural language processing \cite{Joshi1991}, in particular, and
linguistic studies in general.

\vspace{1cm}

The authors are grateful to FAPESP and CNPq (Brazil) and the Human
Frontier Science Program (RGP39/2002) for financial support. Thanks
are also due to several students from NILC for their invaluable help
in the experiment with human judges.

\bibliographystyle{epj}
\bibliography{lucas}

\end{document}